# SINR and Throughput of Dense Cellular Networks with Stretched Exponential Path Loss

Ahmad AlAmmouri, Jeffrey G. Andrews, and François Baccelli

***Abstract*—Distance-based attenuation is a critical aspect of wireless communications. As opposed to the ubiquitous power-law path loss model, this paper proposes a stretched exponential path loss model that is suitable for short-range communication. In this model, the signal power attenuates over a distance $r$ as $e^{-\alpha r^{\beta}}$, where $\alpha, \beta$ are tunable parameters. Using experimental propagation measurements, we show that the proposed model is accurate for short to moderate distances in the range $r \in (5, 300)$ meters and so is a suitable model for dense and ultradense networks. We integrate this path loss model into a downlink cellular network with base stations modeled by a Poisson point process, and derive expressions for the coverage probability, potential throughput, and area spectral efficiency. Although the most general result for coverage probability has a double integral, several special cases are given where the coverage probability has a compact or even closed form. We then show that the potential throughput is maximized for a particular BS density and then collapses to zero for high densities, assuming a fixed SINR threshold. We next prove that the area spectral efficiency, which assumes an adaptive SINR threshold, is non-decreasing with the BS density and converges to a constant for high densities.**

## I. Introduction

Wireless networks are perennially densifying, as new infrastructure is deployed and more end-devices join the network. This densification not only reduces the nominal transmitter-receiver link distance, but also allows spatial reuse of spectrum through cell splitting. These densification gains have been the key enabler for increasing data rates and handling the ever-increasing wireless data demand [2], [3], and are indispensable for 5G and beyond, as well [4], [5]. As the transmit-receive links become shorter, certain aspects of the propagation are different from propagation in longer links. For example, line-of-sight links become more common, destructive ground bounces are less relevant, and discrete obstructions are a more significant source of attenuation than the comparatively gradual inverse square law experienced by free space radio wave propagation.

In this paper, we carefully study the effects of a stretched exponential path loss model, which is a relatively obscure path loss model that may nevertheless prove to be a good model for short-range propagation in many environments. Specifically, we assume that power attenuates over a link distance $r$ as $\exp(-\alpha r^{\beta})$ where $\alpha$ and $\beta$ are fitting parameters with an intuitive physical meaning that we discuss later. Such a model is fairly general and can be quite accurate over short to medium ranges (less than a few hundred meters), as has been established in [6]–[9]. Further, we show that this model has attractive analytical properties, for example key metrics such as SIR and throughput can be derived in fairly simple forms; including results for ultradense networks. Note that the proposed model captures the exponential path loss model as a special case.

The authors are with the Wireless Networking and Communications Group (WNCG), The University of Texas at Austin, Austin, TX 78701 USA. (Email: {alammouri@utexas.edu, jandrews@ece.utexas.edu, francois.baccelli@austin.utexas.edu}). This work has been supported by the National Science Foundation under grant CCF-1514275, and F. Baccellis work was also supported by award #197982 from the Simons Foundation.

Part of this work was presented at IEEE Asilomar Conference on Signals, Systems, and Computers 2017 [1].

### A. Background and Prior Work

Stochastic geometry [18]–[20] has become widely accepted for the mathematical analysis of wireless networks, in particular for determining key metrics such as the signal-to-interference-plus-noise ratio (SINR) and the per-user rate, both of which are random variables and thus characterized by their distributions. The results in [21] showed that a random spatial model, namely a Poisson point process (PPP), could reasonably capture the irregularity and randomness in current base station (BSs) deployments, while providing a high degree of analytical tractability that had lacked in previous cellular network models. A considerable amount of follow on work has generalized and extended [21], as well-summarized in [22]–[24].

The vast majority of this work, and indeed nearly all work in wireless communications including in industry, uses a power-law propagation model where signal power attenuates over a distance $r$ as $r^{-\eta}$, with $\eta$ being called the path loss exponent. Such a model is also referred to as the Hata model [17], and stems from electromagnetic fundamentals, where in free-space $\eta = 2$. With a power-law model, [21], [25] and other papers have shown that the signal-to-interference ratio (SIR) distribution is independent of the BS density. Intuitively, this means that the interference and desired signal power increase at the exact same pace as the network densifies: this property is referred to as *density invariance* or *SINR invariance*. This implies the network can be densified indefinitely and spatial reuse can be harvested over arbitrarily small cell areas.

However, recent studies – both empirical and theoretical – have indicated that this conclusion is quite a bit too optimistic [26]. In fact, for large but realistic densities that could occur quite soon in urban and indoor scenarios, and for several different network and propagation models, it has been argued that not only the SIR – but also the overall throughput – could collapse and approach zero [14], [27]–[30]. The end of cell-splitting gains is called the *densification plateau*; when the throughput actually descends towards zero with higher density, this can be called a *densification collapse*. A key aspect underpinning all these more pessimistic results is the reform of the path loss model to better reflect short-range propagation relative to the standard path loss of $r^{-\eta}$, which treats all distances $r$ as having the same path loss exponent and thus homogeneous attenuation. Homogeneous attenuation is not observed in most real networks: instead, not only the attenuation but also the *rate* of attenuation increases with distance, due to ground bounces, obstructions, and other effects. Ignoring this fact is a critical mistake when analyzing ultradense networks, where both the desired and the strongest interfering transmitters are usually nearby.

TABLE I: Path loss models for dense networks.

| Ref. | Path loss model | Parameters |
|---|---|---|
| Proposed | $PL_1(r) = Ae^{-\alpha r^{\beta}}$ | $A, \alpha, \beta > 0$ |
| [10] | $PL_2(r) = Ae^{-\alpha r^{\beta}} r^{-\eta}$ | $A, \alpha, \beta, \eta > 0$ |
| [11] | $PL_3(r) = Ae^{-\alpha r} r^{-\eta}$ | $A, \alpha, \eta > 0$ |
| [6]–[8], [12] | $PL_4(r) = Ae^{-\alpha r} r^{-2}$ | $A, \alpha > 0$ |
| [9], [13] | $PL_5(r) = A\min\left(r^{-2}, e^{-\alpha r} r^{-2}\right)$ | $A, \alpha > 0$ |
| [14], [15] | $PL_6(r) = A \sum_{i=1}^{N} r^{-\eta_i} 1_{\{R_i < r \leq R_{i+1}\}}(r)$ | $A > 0, \eta_{i+1} \geq \eta_i \geq 0, R_{i+1} \geq R_i \geq 0$ |
| [16] | $PL_7(r) = A(1 + r)^{-\eta}$ | $A, \eta > 0$ |
| [16] | $PL_8(r) = A(1 + r^{\eta})^{-1}$ | $A, \eta > 0$ |
| [17] | $PL_9(r) = Ar^{-\eta}$ | $A, \eta > 0$ |

Several simple path loss models that might be suitable for dense networks are summarized in Table I. In [14], the multi-slope path loss model is analyzed, where different distance ranges are subject to different path loss exponents. This model is also used in 3GPP-LTE standardization [31] and verified by measurements as in [32]. The results in [14] show that a *densification plateau* or even collapse can exist when the value of the short range path loss exponent $\eta_1$ is sufficiently small, namely $\eta_1 \leq 1$. The models used in [16] are closely related to [14], with only two slopes.

Moving away from power-law path loss, the papers [6]–[13] consider an exponential component in the path loss. The models in [7]–[9] are based on actual measurements, but mostly in an indoor environment. However, the authors in [6] derived a similar path loss model using a simple stochastic model based on the theory of random walks in small urban cells. This model is also closely connected to the empirically supported multi-slope model of [14], and in fact is a limiting case of that model with a large number of slopes, as seen in Fig. 1. The exponential model has received little attention to date for system-level performance evaluation, but there are a few other relevant works. It was incorporated into a stochastic geometry analysis in [12], which also included power-law path loss, and the resulting complexity resulted in only a preliminary analysis and ultimately Chernoff bounds on the SIR distribution. A special case of our model with $\beta = 1$ was used in [33] to evaluate the performance of an ad hoc network. A final interesting result is [10], where the authors derived the path loss equation based on ray propagation in random lattices as a model for radio propagation in urban areas. To the best of the authors' knowledge, a comprehensive analysis of the effect of exponential path loss on SIR and throughput has never been performed prior to this paper.

### B. Contributions and Organization

We propose a path loss model given by $e^{-\alpha r^{\beta}}$, where $\alpha, \beta > 0$ are tunable parameters. We verify this model with the same measurements used in [6], and show that it provides the best fit of any of the candidate models in Table I. We use the proposed model to study network coverage probability, or equivalently the SIR distribution. We prove the SIR coverage probability dependence on the BS density and show that the coverage probability approaches zero for high BS densities.

We then turn our attention to the achievable rate. We consider two rate metrics: the *potential throughput* and the average *area spectral efficiency* (ASE). The potential throughput, as in [14], [34], is defined as $\mathcal{R}(\lambda, \theta) = \lambda \log_2(1 + \theta)\mathbb{P}\{\text{SIR} \geq \theta\}$, which implicitly assumes a fixed rate transmission from all BSs in the network, and has units of bps/Hz/m$^2$. The potential throughput is optimistic in that it assumes all BSs have a full buffer and can transmit (thus the word "potential"), but is pessimistic in that a given node's rate cannot be raised when $\text{SIR} \geq \theta$, resulting in wasted throughput; nor can it be lowered when $\text{SIR} < \theta$, resulting in unnecessary outages. We prove that the potential throughput is maximized for a finite value of the BS density then it collapses and reaches zero for high density.

The average ASE, defined as $\mathcal{E}(\lambda) = \mathbb{E}\left[\lambda \log_2\left(1 + \text{SIR}\right)\right]$, also assumes full buffers and has units of bps/Hz/m$^2$, but it allows each link to adapt its rate to the optimum value (the Shannon's rate), for a given SIR, thus avoiding outages at low SIR and the wasting of rate at high SIR. Therefore the average ASE upper bounds the potential throughput while still being a realistic measure, since modern networks allow rapid link adaptation. We derive a general expression for the ASE, prove that it is always finite and that it is a non-decreasing function of the density. Moreover, we derive a closed-form expression for its limit when the BS density approaches infinity, which is a simple finite constant. Thus, there is no densification collapse in terms of the ASE, but there is a densification plateau.

Note that the ASE saturation is significantly different from [21] under power-law path loss where the ASE scales sub-linearly with the BS density. In fact, it was also reported in [34]–[36] that, under LoS/NLoS communications, the ASE does not linearly increase with the BS density and an optimal density might exist. In [35] ( [36]), the authors considered a general multi (dual)-slope path loss model that accounts for LoS/NLoS and observed numerically that the ASE may have different behaviors with respect to the BS density depending on the considered density and path loss parameters. However, the limiting case where the BS density approaches infinity was not analyzed in [34]–[36].

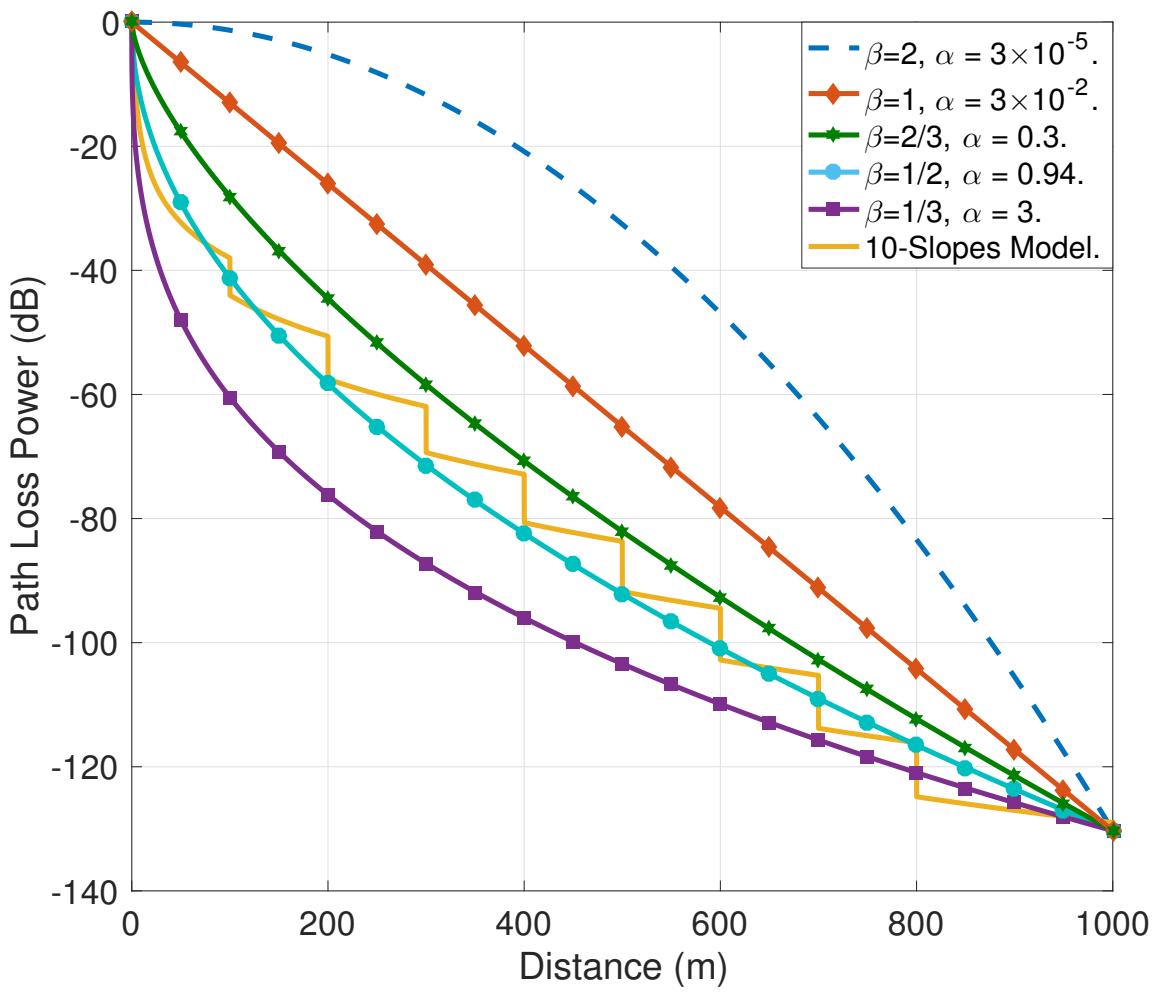

Fig. 1: Path loss power vs. distance for the stretched exponential and 10-slope power-law path loss models.

The rest of the paper is organized as follows. In Section II, we present the system model and methodology of the analysis. In Section III, we verify the proposed model with actual measurements. Section IV is the main technical section of the paper, where we derive expressions for the considered performance metrics. Numerical and simulation results with discussion are presented in Section V before the conclusion in Section VI. Most proofs are in the Appendix to enhance readability.

## II. System Model

A downlink cellular network is considered where the BSs are spatially distributed according to a homogeneous Poisson point process (PPP) $\Psi$ with density $\lambda$ [19]. Closest BS association is assumed where each BS serves exactly one user whose location is uniformly distributed in the BS association area. All BSs transmit with a fixed power, and all BSs and users are equipped with a single antenna. All signals are subject to Rayleigh fading, with a unit mean exponentially distributed small-scale fading power, denoted as $h$. All channels are assumed to be i.i.d. and independent from the BS and user locations. Other fading models, such as shadowed $\kappa-\mu$, $\eta-\mu$, and two rays plus defuse power (TWDP) can also be used. However, it was shown in [37], [38] that using these models with the power-law path loss model $r^{-\eta}$ does not help in capturing the densification plateau and the density-invariance property still holds. Hence, we focus on the more tractable Rayleigh fading model since the key factor to capture the plateau is the path loss model.

The large-scale attenuation is captured by the proposed path loss model, where the path loss attenuation over a distance $r$ is given by $\exp(-\alpha r^{\beta})$, where $\alpha, \beta > 0$ are tunable parameters. For $\beta = 1$, the proposed model captures the exponential path loss model used in [33]. As mentioned earlier, this model is a limiting case of the multi-slope model [14] with many slopes as illustrated in Fig. 1. As Fig. 1 shows[1], the two tunable parameters provide flexibility to capture different propagation behaviors. Note that the value of $\alpha$ is chosen such that the path loss power is fixed at a distance 1 km for all the models to allow us to compare the different propagation scenarios. Otherwise, if the value of $\alpha$ is fixed for all values of $\beta$, the path loss power will increase very fast with the distance when $\beta = 2$, which makes comparing it with the other models difficult.

The proposed model also describes the case where the attenuation is mostly due to obstructing objects, each of which causes a multiplicative attenuation of $\alpha$ (on average), with the number of total obstacles in the path scaling as $r^{\beta}$. This scaling can be physically interpreted for different values of $\beta$ as follows:

- $\beta = 1$: In this case, the number of obstacles scales linearly with the distance between the user and the BS. It was shown in [39] using random shape theory that if the obstacles (buildings) are uniformly distributed in the region with random orientations, the number of obstacles crossing the line connecting the BS and the user scales linearly with the distance. In this case, $\alpha$ is a function of the attenuation of each block.
- $\beta = 2$: Using a similar argument as for $\beta = 1$, the number of obstacles scales with $r^2$ if we allow the signal to propagate within a sector of the disk centered at the user location and extending to the BS. Hence, the attenuation will be a function of the number of obstacles located in this area which is a function of $r^2$. In this case, $\alpha$ is a function of the attenuation of each block, the section's angle, and the way the signal moves within the section.
- $1 < \beta < 2$: This region does not have a precise physical interpretation as in the cases of $\beta = 1$ and $\beta = 2$. However, it can be observed empirically if the number of obstacles scales faster than $r$ but slower than $r^2$.
- $\beta < 1$: This range agrees with the results shown in [10], which was based on ray propagation in percolating lattices as a model for the signal propagation in urban areas with regular building blocks. The parameter $\alpha$ in this case depends on the reflectivity of the obstacles and the properties of the considered lattice. Note that the extra term in [10] ($r^{-\eta}$) is relevant for the extremely small distance ranges that are not typical in the cellular network scenario. More discussion on this point is provided in Section III.

Mathematically, the proposed model is continuous and bounded, which helps in incorporating it in stochastic geometry analysis. Moreover, since $\exp(-\alpha r^{\beta}) < 1$ for all $r > 0$, it has an advantage over the standard power-law path loss model of $r^{-\eta}$ which is not suitable for small distances since $r^{-\eta} > 1$

[1]In the 10-slope model, for distances less than 1m, $\eta = 0$, then it is assumed to be 1.9 and increases by 0.3 each 100 m.

for $r < 1$ [15] [2].

Overall, the advantages of the proposed model can be summarized as:

1) It is suitable for short distances and it provides the best fit to the available measurements compared to the other path loss models (shown in Table I) as will be shown in Section III.
2) The ability to derive relatively simple expressions and prove informative insights on the network performance, especially in the limiting case of a very dense network scenario.
3) Flexibility; since it has two tunable parameters that allow it to capture different propagation behaviors as shown in Fig. 1.
4) It can be physical interpreted for some values of $\beta$ as discussed previously in this section.
5) It is practically appropriate since it is bounded and continuous.

However, it has also some drawbacks; it is not suitable for long distances and it does not capture accepted and well-established path loss models as special cases (e.g. power-law path loss). Hence, we neither claim that the stretched exponential model is the best path loss model nor that it can replace existing path loss models. Rather, it is an alternative model that is suitable for short distance communications and it allows us to prove interesting properties that have been observed in practice.

Moving to the performance evaluation, we focus on three metrics: *coverage probability*, *potential throughput*, and *average area spectral efficiency*. The *coverage probability* is defined as $P_{\mathrm{cov}}(\lambda,\theta) = \mathbb{P}\left(\mathrm{SIR} \geq \theta\right)$, where SIR is the signal to interference ratio and $\theta$ is the coverage threshold. The second performance metric is the *potential throughout* which is defined as in [14]

$$\mathcal{R}(\lambda,\theta) = \lambda \log_2(1+\theta)\mathbb{P}\left\{\mathrm{SIR} \geq \theta\right\}, \tag{1}$$

where $\lambda$ is the BS density. The potential throughput can be used to study the throughput scaling with the BS density assuming that all the users have the same rate $\log_2(1+\theta)$ bps/Hz, and the transmission is considered successful only if $\mathbb{P}\left\{\mathrm{SIR} \geq \theta\right\}$ [14], [34]. So $\theta$ is the minimum SIR threshold required to decode the transmit messages successfully. This performance metric has units in bps/Hz/m$^2$ and it represents the maximum number of successful transmissions possible per unit area.

Although the potential throughput can provide valuable insights on the network performance as in [14], [34], it can be argued that it is pessimistic, since users cannot exploit SIR values higher than the threshold $\theta$ or lower their rates when $\mathrm{SIR} < \theta$ to avoid outage. Hence, the threshold $\theta$ can be optimized to enhance the network performance. One way to do that is to choose $\theta$ such that $\theta^* = \underset{\theta \in \mathbb{R}^+}{\mathrm{argmax}}\ R(\lambda,\theta)$, in this case the average potential throughput is maximized, but not the instantaneous throughput. To optimize the instantaneous throughput, the threshold $\theta$ should be set to its maximum value instantaneously such that the condition $\mathrm{SIR} \geq \theta$ is satisfied, which means that $\theta = \mathrm{SIR}$. The corresponding maximum instantaneous throughput in this case is $\lambda \log_2(1+\mathrm{SIR})$, and on average it is given by

$$\mathcal{E}(\lambda) = \lambda \mathbb{E}\left[\log_2\left(1+\mathrm{SIR}\right)\right], \tag{2}$$

where $\mathcal{E}(\lambda)$ still has units in bps/Hz/m$^2$ and is commonly called by the *average area spectral efficiency (ASE)* [34], [40][3]. However, it is also an optimistic measure since it assumes that the Shannon's rate is achievable and it requires an instantaneous update of the used transmission codes according to the SIR values. Hence, we study both the potential throughput and the ASE, since the actual network performance lies in between. Other performance metrics, such as the bit-error probability, can be found using the coverage probability (SINR distribution) following the methods mentioned in [22].

As a final remark before delving into the analysis, we want to emphasize that we assume a saturated network where each BS serves exactly one UE in any given resource block. This implicitly means that the UE density scales with the BS density and no BSs are idle. Otherwise, the analysis should be adapted as in [29], [34].

## III. Fitting the path loss model

In this section, we verify our proposed path loss model with actual measurements. We use the measurements provided in [6], based on an experiment in Rome, Italy, by Ericsson. For a detailed description of the experimental setup and measurements location, refer to [6, Section VII]. The proposed model is compared with the models provided in the first column of Table II, where the parameter $A$ is a constant that captures the effects of the antenna gains, transmit power, and wavelength.

To quantify the difference between the proposed model and the measurements, we use the RMS error (in dB) between the measured data and the proposed models similar to [6]. The RMS error for the $i^{\mathrm{th}}$ model is defined as

$$\sigma_i = \sqrt{\frac{\sum_{j=1}^{n}\left(PL_{\mathrm{m}}(r_j) - PL_i(r_j)\right)^2}{n}}. \tag{3}$$

All the parameters in the chosen path loss models are found by numerical evaluation of least square problems and are shown in Table II along with the associated RMS error values. Before presenting the results, we want to emphasize the following points:

1) The measurements are collected in the distance range $r \in [5, 315]$ m. Hence, it is suitable for small or medium sized cells in urban environments. More discussion on this point is given in Section V.

[2]Although by adding the parameter $A$ as in Table I, the value of $r$ for which the path loss exceeds 1 can be adjusted, it still suffers from the singularity at $r \to 0$, which can produce undesirable artifacts when analyzing the network performance.

[3]Note that the given interpretation of the definition of $\mathcal{E}(\lambda)$ assumes that the channel is constant over each transmitted codeword. However, this quantity serves as an upper bound on the achievable rate if the transmitted messages suffer from multiple small-scale fluctuations during each codeword as indicated in [41].

TABLE II: Parameter values with the associated RMS error.

| Model | RMS Error (dB) | Parameter 1 | Parameter 2 | Parameter 3 | Parameter 4 |
|---|---|---|---|---|---|
| $PL_1(r) = Ae^{-\alpha r^{\beta}}$ | 3.5072 | $A = 0.0094$ | $\alpha = 0.9019$ | $\beta = 0.5210$ | - |
| $PL_2(r) = Ae^{-\alpha r^{\beta}} r^{-\eta}$ | 3.5072 | $A = 0.0094$ | $\alpha = 0.9019$ | $\beta = 0.5210$ | $\eta = 0$ |
| $PL_3(r) = Ae^{-\alpha r} r^{-\eta}$ | 3.7429 | $A = 0.0615$ | $\alpha = 0.0286$ | $\eta = 1.9391$ | - |
| $PL_4(r) = Ae^{-\alpha r} r^{-2}$ | 3.7442 | $A = 0.0758$ | $\alpha = 0.0281$ | - | - |
| $PL_5(r) = A \min\left(e^{-\alpha r} r^{-2}, r^{-2}\right)$ | 3.7442 | $A = 0.0758$ | $\alpha = 0.0281$ | - | - |
| $PL_6(r) = \sum_{i=1}^{2} r^{-\eta_i} 1_{\{R_i < r \leq R_{i+1}\}}(r)$ | 4.6682 | $A = 5.3576$ | $\eta_1 = 3.4892$ | $\eta_2 = 4.0345$ | $R_1 = 142.7778$ |
| $PL_7(r) = A(1+r)^{-\eta}$ | 5.8695 | $A = 561.07$ | $\eta = 4.7729$ | - | - |
| $PL_8(r) = A(1+r^{\eta})^{-1}$ | 6.0778 | $A = 286.11$ | $\eta = 4.6466$ | - | - |
| $PL_9(r) = Ar^{-\eta}$ | 6.0779 | $A = 286.16$ | $\eta = 4.6467$ | - | - |

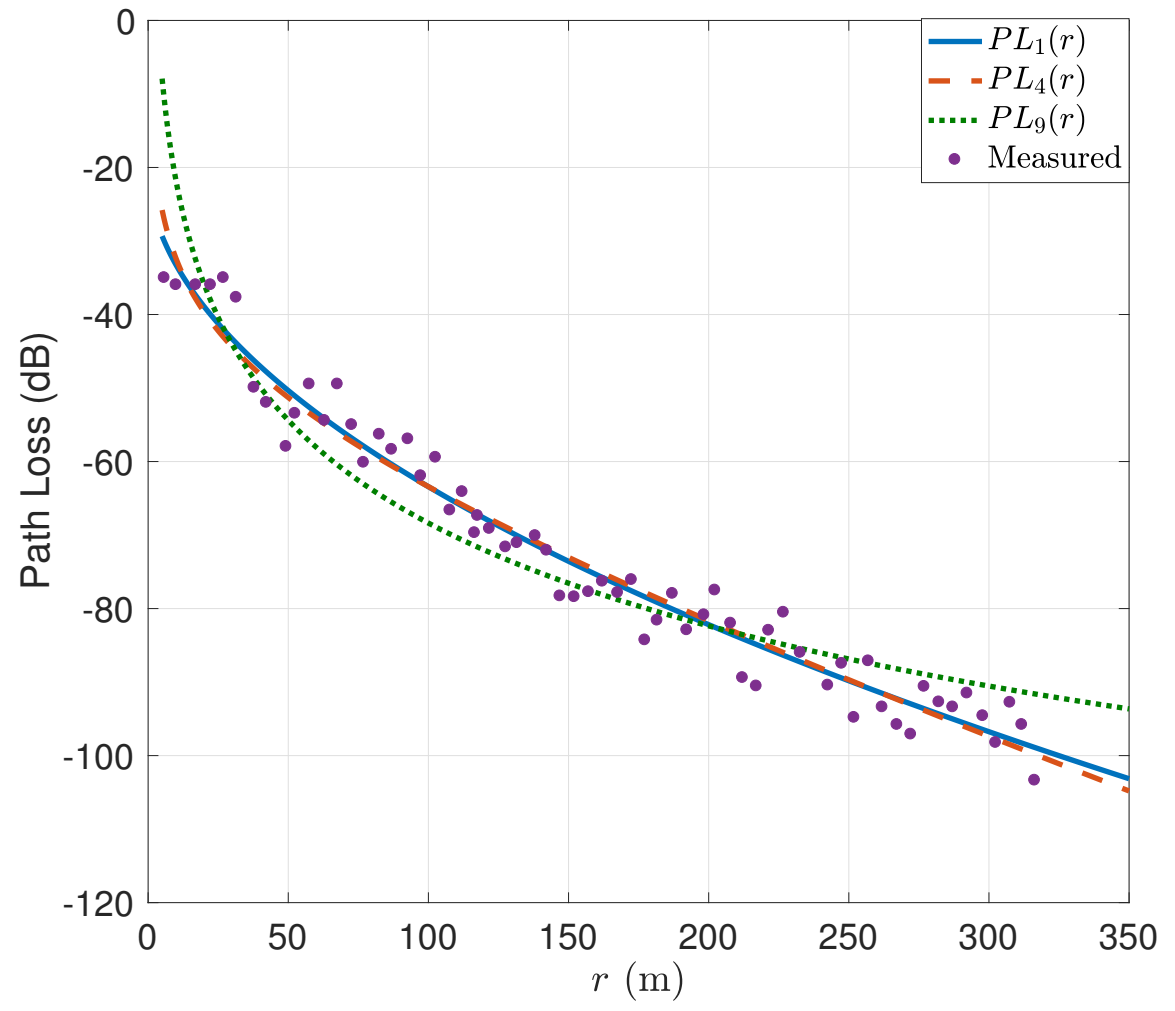

Fig. 2: The measured and proposed path loss models. Our model is denoted by $PL_1(\cdot)$.

2) Distances less than 5 meters are not included in the analysis. However, this distance range is not typical in cellular networks, even in the very dense networks, due to the common difference in elevation between the users and the BSs [28].
3) The mean-square optimal path loss model (optimal non-increasing function) has an RMS error value $\sigma_{\text{opt}} = 2.04$ dB according to [6].

As Table II shows, our proposed model provides the best fit, in the mean square sense, among all considered path loss models. Moreover, it shows that there is a relatively big difference between it and the standard power-law model $PL_9(r)$, which means that standard power-law path loss is not suitable for small distances. Note that the difference between the models $PL_4(r)$ and $PL_5(r)$ is mainly in the small distances range, $r \lesssim 1$. Hence there is no observable difference between them in the considered range. The same argument applies to the models $\{PL_8(r), PL_9(r)\}$ and $\{PL_1(r), PL_2(r)\}$.

To visualize the differences between the different models, Fig. 2 compares the measured data, our proposed model, and the models $PL_4(r)$ and $PL_9(r)$. As Fig. 2 shows, there is a little difference between our model and $PL_4(r)$. However, there is a notable difference between our model and $PL_9(r)$. In the next section, we will be more interested in the values of $\beta$ that are in the form $\beta = 2/(n+1)$ for any non-negative integer $n$. However, if we add this constraint on $\beta$, the optimal values for the proposed model are $\alpha = 1.037$ and $\beta = 0.5$ $(n = 3)$ with RMS error $\sigma = 3.51$ dB, which is still better than the other models.

Although the results in Table II and Fig. 2 confirm the flexibility of the stretched exponential function to model the path loss over the distance range $r \in [5, 350]$ in urban areas, we do not claim that it is suitable for longer distances. More specifically, the power-law path loss will dominate for long distances and will be a more accurate model than the exponential one. However, for dense cellular networks, the interference is dominated by BSs in a certain range and not affected by BSs far away. In Section V, we verify by simulation that characterizing the path loss in the range $r \in [5, 350]$ is sufficient to study the performance of dense cellular networks.

## IV. Performance Analysis

In this section, we derive expressions for the considered performance metrics and study the effect of densification on these metrics. We start each subsection by presenting the analytical expressions then we use the derived formulas to show how it scales with the BS density. We focus on the interference-limited network since it is the typical case for dense cellular networks, but we also highlight the results that can be extended to include noise in Section IV-D.

### A. Coverage Probability

We consider a user located at the origin, without loss of generality, and the mean performance then represents the average performance for all users in the network [19]. Let the aggregate network interference be denoted by $\mathcal{I}$. Then based on our system model, the coverage probability is given by

$$\mathbb{P}\left\{\text{SIR} \geq \theta\right\} = \mathbb{P}\left\{\frac{h_o e^{-\alpha r^{\beta}}}{\mathcal{I}} \geq \theta\right\} = \mathbb{E}_r\left[\mathcal{L}_{\mathcal{I}}\left(\theta e^{\alpha r^{\beta}}\right)\right], \tag{4}$$

where $h_o$ is the fading power of the desired signal, and $r$ is the serving distance (the distance between the user and its serving BS). Equation (4) follows by exploiting the exponential distribution of $h_o$ similar to [21] and $\mathcal{L}_{\mathcal{I}}(\cdot)$ is the Laplace transform (LT) of the aggregate interference PDF. The expectation in (4) is with respect to (w.r.t.) the serving distance, which has a PDF given by $f_R(r) = 2\pi\lambda r \exp(-\pi\lambda r^2)$ [21].

In the next theorem, we present the most general expression for the coverage probability.

**Theorem 1.** *The downlink coverage probability of a cellular network with BS density $\lambda$ and a stretched exponential path loss model $PL_1(r) = Ae^{-\alpha r^{\beta}}$ is given by*

$$P_{\rm cov}(\lambda,\theta) = 2\pi\lambda \int_0^{\infty} e^{-\pi\lambda r^2} \mathcal{L}_{\mathcal{I}}\left(\theta e^{\alpha r^{\beta}}\right) r dr, \tag{5}$$

*where $\mathcal{L}_{\mathcal{I}}(\cdot)$ is the Laplace transform of the aggregate interference PDF and given by*

$$\mathcal{L}_{\mathcal{I}}(s) = \exp\left(-\frac{2\pi\lambda}{\beta\alpha^{\frac{2}{\beta}}} \int_0^{e^{-\alpha r^{\beta}}} \frac{s}{1+sy}\left(-\ln(y)\right)^{\frac{2-\beta}{\beta}} dy\right). \tag{6}$$

*Proof.* Refer to Appendix A. □

In pursuit of simpler results, the coverage probability for some special values of $\beta$ are presented in the following Corollaries.

**Corollary 1.** *For the special case of $\beta = \frac{2}{1+n}$, where $n$ is any non-negative integer ($n \in \mathbb{Z}^+$), the coverage probability in* (5) *reduces to*

$$P_{\rm cov}(\lambda,\theta) = 2\pi\lambda \int_0^{\infty} r \exp\left(\sum_{k=0}^{n+1} \lambda\, a_k(\theta)\, r^{\frac{2k}{n+1}}\right) dr, \tag{7}$$

*where,*

$$a_k(\theta) = \begin{cases} \frac{\pi(n+1)!}{k!\alpha^{n-k+1}} \operatorname{Li}_{(n-k+1)}(-\theta), & 0 \le k \le n. \\ -\pi, & k = n+1. \end{cases} \tag{8}$$

*and $\operatorname{Li}_{(k)}(\cdot)$ is the $k^{th}$ order polylogarithmic function [42], which can be represented as*

$$\operatorname{Li}_{(n-k+1)}(\theta) = \frac{\theta}{\Gamma[n-k+1]} \int_0^{\infty} \frac{x^{n-k}}{e^x - \theta} dx. \tag{9}$$

*Proof.* Refer to Appendix B. □

**Corollary 2.** *For the special case of $\beta = 1$, the coverage probability in* (5) *reduces to*

$$P_{\rm cov}(\lambda,\theta) = \exp\left(\frac{2\pi\lambda}{\alpha^2} \operatorname{Li}_2(-\theta)\right) \times \left(1 - \frac{2\pi\sqrt{\lambda}}{\alpha} y(\theta) e^{\frac{\pi\lambda y(\theta)^2}{\alpha^2}} \operatorname{Q}\left(\frac{\sqrt{2\pi\lambda}}{\alpha} y(\theta)\right)\right), \tag{10}$$

*where $y(\theta) = \ln(1+\theta)$ and $\operatorname{Q}(\cdot)$ is the Q-function [42].*

*Proof.* Follows from Corollary 1 given that $\operatorname{Li}_1(-\theta) = -\ln(1+\theta)$ and then by using [43, eq. 5.12.20] after simple mathematical manipulations. □

**Corollary 3.** *For the special case of $\beta = 2$, the coverage probability in* (5) *reduces to*

$$P_{\rm cov}(\lambda,\theta) = \exp\left(-\frac{\pi\lambda}{\alpha}\ln(1+\theta)\right). \tag{11}$$

*Proof.* Follows from Corollary 1 given that $\operatorname{Li}_1(-\theta) = -\ln(1+\theta)$. □

Corollaries 1, 2, and 3 provide exact expressions for the coverage probability. However, the expression is more complicated in Corollary 1. Hence, we provide simpler bounds in the next Corollary for the case of $\beta = 2/(n+1)$.

**Corollary 4.** *The coverage probability in* (7) *is upper-bounded by*

$$P_{\rm cov}(\lambda,\theta) \le \exp\left(\lambda a_0(\theta)\right) \Gamma^{'}\left(1, 0, -\frac{\lambda^{1-\frac{n}{n+1}} a_n(\theta)}{\pi^{\frac{n}{n+1}}}, \frac{-n}{n+1}\right), \tag{12}$$

$$\le \exp\left(\lambda \frac{\pi(n+1)!}{\alpha^{n+1}} \operatorname{Li}_{(n+1)}(-\theta)\right), \tag{13}$$

$$\le \exp\left(-\lambda \frac{\pi(n+1)!}{\alpha^{n+1}} \ln(1+\theta)\right), \tag{14}$$

*with equality in* (12) *for $n = 1$ ($\beta = 1$) and in* (13) *and* (14) *for $n = 0$ ($\beta = 2$). Here $\Gamma^{'}(\cdot,\cdot,\cdot,\cdot)$ is the extended incomplete Gamma function [44]. Moreover, the coverage probability is lower bounded by*

$$P_{\rm cov}(\lambda,\theta) \ge \exp\left(\sum_{k=0}^{n} a_k(\theta) \lambda^{1-\frac{k}{n+1}} \pi^{\frac{-k}{n+1}} \Gamma\left(1 + \frac{k}{n+1}\right)\right), \tag{15}$$

*with equality when $n = 0$ ($\beta = 2$), and $\Gamma(\cdot)$ is the Gamma function [44].*

*Proof.* Refer to Appendix C. □

Next, we study the effect of densification on the coverage probability. Starting from the simplest expression in Corollary 3 for $\beta = 2$, the coverage probability decreases with increasing $\lambda$, which means that the increase in the aggregate network interference dominates the decrease in the serving distance between each user and its serving BS. Moreover, the coverage probability approaches zero when $\lambda \to \infty$. For a general value of $\beta$, the effect of increasing $\lambda$ is not clear from the exact expression in Corollary 1. However, we can exploit the derived bounds to provide insights on the effect of network densification on the coverage probability as in the following Corollary.

**Corollary 5.** *For $\beta = \frac{2}{n+1}$, where $n$ is any non-negative integer, the coverage probability approaches zero when $\lambda \to \infty$.*

*Proof.* Follows from the upper bound in (14). □

Note that the obtained results show that the density-invariance property does not hold for the stretched exponential path loss model, unlike the standard power-law path loss model [21].

### B. Potential Throughput

The potential throughput can be directly obtained from the coverage probability expressions using (1), hence we will only provide expressions for it when it is necessary. The potential throughput for $\beta = 2$ is given by

$$\mathcal{R}(\lambda, \theta) = \lambda \exp\left(-\frac{\pi\lambda}{\alpha}\ln(1+\theta)\right)\log_2(1+\theta). \tag{16}$$

Based on (16), the potential throughput also approaches zero when $\lambda \to \infty$. Furthermore, the potential throughput is a log-concave function of $\lambda$ and has a unique maximizer given by $\lambda^* = \frac{\alpha}{\pi \ln(1+\theta)}$. After this value, the throughput collapses and approaches zero. Hence, there are no benefits from network densification after this point. Unfortunately, there is no closed form expression for the maximizer in the general case given in Corollary 1. Nevertheless, we can also exploit the compact form obtained and the derived bounds to obtain insights about the throughput scaling with the BS density. These insights are summarized in the following Corollary.

**Corollary 6.** *For $\beta = \frac{2}{n+1}$, where $n$ is any non-negative integer, the potential throughput is a log-concave function of the BS density $\lambda$ with a finite maximizer $\lambda^*$. This maximizer is given by $\lambda^* = \frac{\alpha}{\pi \ln(1+\theta)}$ for the special case of $\beta = 2$. Moreover, the potential throughput approaches zero when $\lambda \to \infty$.*

*Proof.* Refer to Appendix D. □

The last corollary also highlights another major difference between the proposed model and the power-law path loss model, where the potential throughput increases at least linearly with the BS density [21]. It also agrees with the findings in [16] where the authors used bounded path loss models in the form of $\mathrm{PL}_7(\cdot)$ and $\mathrm{PL}_8(\cdot)$ shown in Table I. They found that the potential throughput scales as $\lambda e^{-K\lambda}$ with the BS density, where $K$ is a positive constant. This scaling law has exactly the same form as in the case of $\beta = 2$ shown in (16). For a general $\beta$, both of the scaling laws, $\lambda e^{-K\lambda}$ and the one in the last corollary, are log-concave functions with unique finite maximizers.

### C. Area Spectral Efficiency

As shown in [21], the ASE can be expressed in terms of the coverage probability as follows

$$\begin{aligned}\mathcal{E}(\lambda) &= \mathbb{E}\left[\lambda \log_2\left(1+\mathrm{SIR}\right)\right],\\ &= \lambda \log_2(e)\int_0^\infty \mathbb{P}\left\{\ln(1+\mathrm{SIR}) > t\right\}dt, \quad (17)\\ &= \lambda \log_2(e)\int_0^\infty \frac{P_{\mathrm{cov}}(t)}{t+1}dt, \quad (18)\end{aligned}$$

where (17) follows from the fact that the SIR is a non-negative random variable and (18) by a simple change of variables. Using (18), we can use the bounds and expressions that we already derived for the coverage probability to find the ASE as in the following theorem and corollary.

**Theorem 2.** *The area spectral efficiency under the proposed stretched exponential model is given by*

$$\mathcal{E}(\lambda) = 2\pi\lambda^2 \log_2(e)\int_0^\infty\int_0^\infty e^{-\pi\lambda r^2}\mathcal{L}_{\mathcal{I}}\left(te^{\alpha r^\beta}\right) rdr\frac{dt}{t+1}, \tag{19}$$

*where $\mathcal{L}_{\mathcal{I}}(\cdot)$ is given in (6). For the special case of $\beta = \frac{2}{n+1}$, where $n$ is any non-negative integer, the ASE simplifies to*

$$\mathcal{E}(\lambda) = 2\pi\lambda^2 \log_2(e)\int_0^\infty\int_0^\infty r\exp\left(\sum_{k=0}^{n+1}\lambda\ a_k(t)\ r^{\frac{2k}{n+1}}\right)\frac{drdt}{t+1}, \tag{20}$$

*where $a_k(\cdot)$ is given in (8).*

*Proof.* Follows by substituting the coverage probability in (18) by the expressions given in Theorem 1 and Corollary 1, respectively. □

**Corollary 7.** *For the special case of $\beta = 2$, the ASE in (19) simplifies to the constant*

$$\mathcal{E}(\lambda) = \log_2(e)\frac{\alpha}{\pi}\ \forall\lambda > 0. \tag{21}$$

*Proof.* Follows directly by substituting (11) in (18) and evaluating the integral. □

Corollary 7 shows that multiplying by $\lambda$ in (2), to account for the number of links per unit area, fully balances the decrease in the SIR distribution w.r.t. $\lambda$. This means that densifying the network will not result in any gains in terms of the ASE. For a general value of $\beta$, we use the derived bounds for the coverage probability to bound the ASE as in the next Corollary.

**Corollary 8.** *The average area spectral efficiency, for the case of $\beta = 2/(n+1)$ where $n$ is any non-negative integer, is a log-concave function w.r.t. $\lambda$ and it is upper bounded by a constant*

$$\mathcal{E}(\lambda) \leq \frac{\alpha^{n+1}}{\pi(n+1)!}\log_2 e, \tag{22}$$

*with equality for $n = 0$ ($\beta = 2$).*

*Proof.* Follows directly by substituting the bound (14) in (18) and evaluating the integral. The proof of log-concavity is similar to the proof of Corollary 6. □

The last corollary shows that the ASE is upper bounded by a constant that is independent of $\lambda$ and it is also a log-concave function. Which implies that the limit of the ASE exists and finite when $\lambda \to \infty$. Interestingly, the ultradense limit is the same as the upper bound provided in the last corollary.

**Theorem 3.** *The ASE is a non-decreasing function of $\lambda$ and*

$$\lim_{\lambda\to\infty}\mathcal{E}(\lambda) = \frac{\alpha^{n+1}}{\pi(n+1)!}\log_2 e. \tag{23}$$

*Proof.* Refer to Appendix E. □

Since the potential throughput, which is a pessimistic measure, approaches zero for $\lambda \to \infty$ and the ASE, which is a more optimistic measure, approaches a finite constant for $\lambda \to \infty$, the actual network performance lies in between. This proves the existence of a densification plateau where the network rate, in the best-case scenario, will saturate to a constant. Note that we cannot directly compare the results for the ASE with [14], since the potential throughput was considered there, and analysis for the ASE with multi-slope path loss is currently missing from the literature.

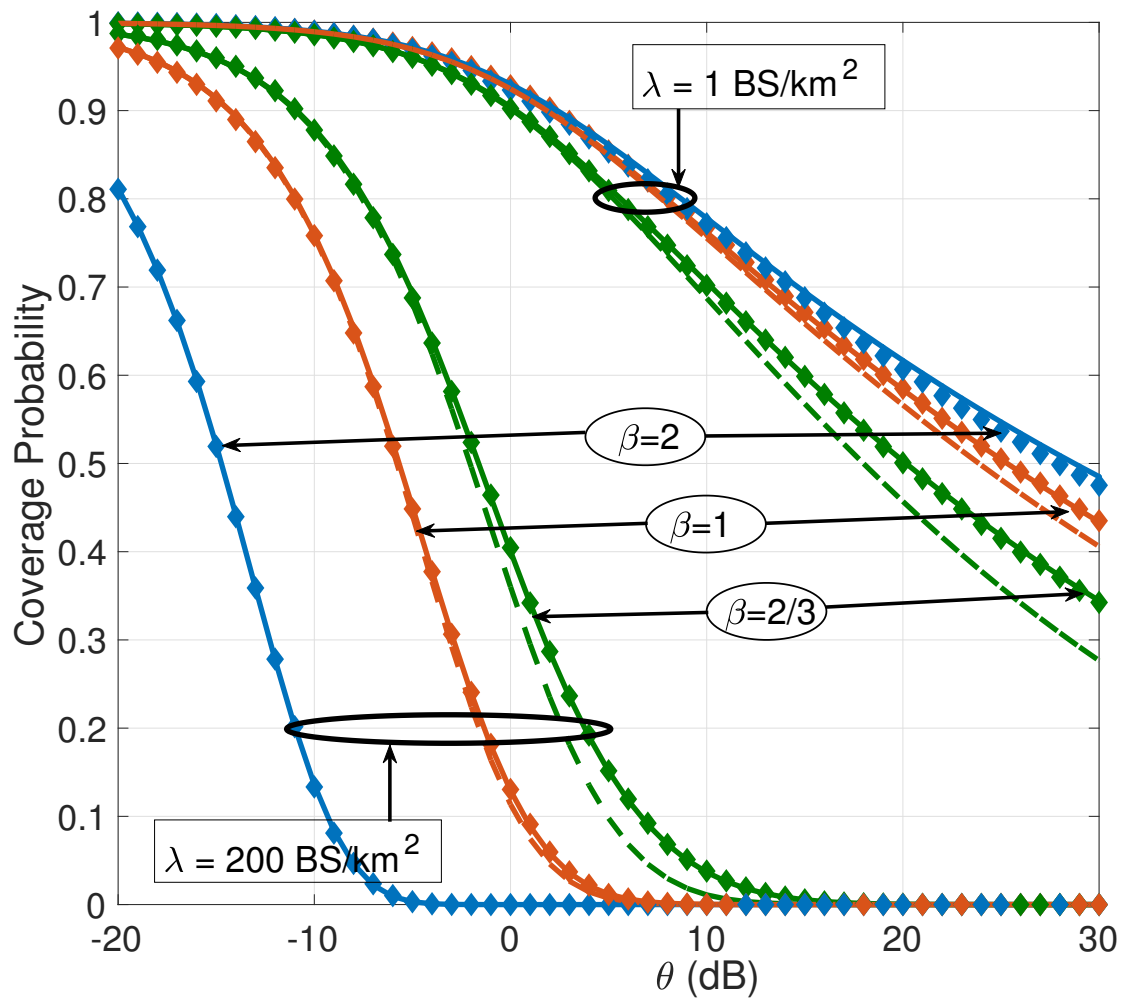

Fig. 3: The coverage probability vs. $\theta$. The solid, dashed, and diamonds lines represent the exact, the lower bound (15), and the simulation results, respectively.

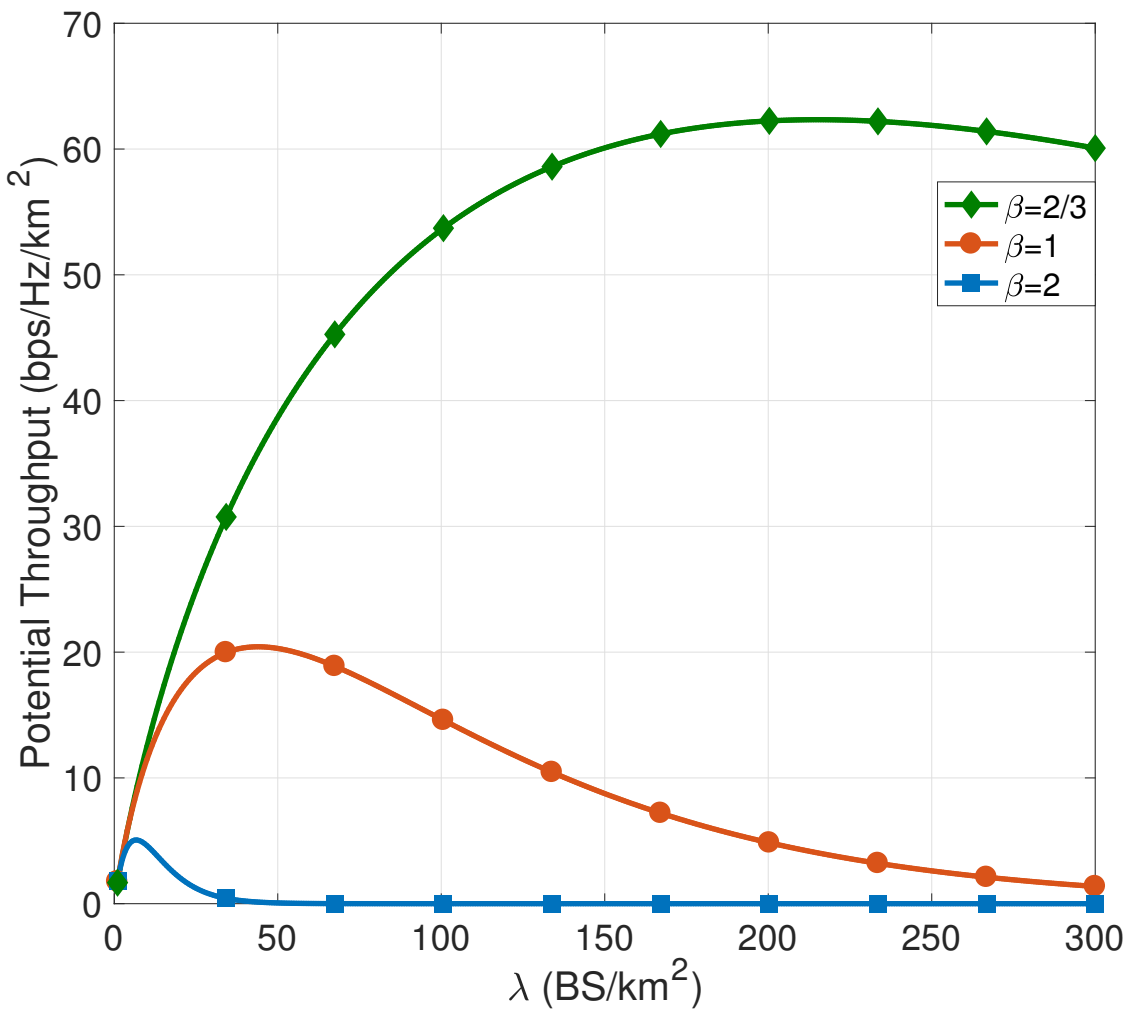

Fig. 4: The potential throughput vs. the BS density ($\lambda$ BSs/km$^2$) for coverage threshold $\theta = 5$ dB. Recall $\beta = 1$ corresponds to standard blocking models in 1-D [39].

### D. SINR Analysis

We have neglected the noise so far and considered an interference-limited network. This assumption is justified for dense cellular networks since the performance is limited by the aggregate network interference that dominates the noise. However, it is straightforward to generalize the analysis by including noise, but at the expense of losing the compact forms in the previous corollaries. Nevertheless, the coverage probability, the potential throughput, and the ASE for the noiseless case are upper bounds for the general case that includes the noise. Hence, we still can draw insights about the general case; these insights are summarized in the following Theorem.

**Theorem 4.** *For the general case, which includes noise in addition to the network interference and $\beta = 2/(n+1)$, where $n$ is any non-negative integer, the coverage probability approaches zero when $\lambda \to \infty$ and the potential throughput is a log-concave function w.r.t. $\lambda$ and it is maximized for a finite $\lambda$ and it collapses and approaches zero for $\lambda \to \infty$. Moreover, the average area spectral efficiency is upper-bounded by the finite constant $\frac{\alpha^{n+1}}{\pi(n+1)!}\log_2 e$ and approaches it as $\lambda \to \infty$.*

*Proof.* The asymptotic gain follows from the fact that $\mathbb{P}\{\text{SINR} \geq \theta\} \leq \mathbb{P}\{\text{SIR} \geq \theta\}$, and we have already shown that for $\beta = 2/(n+1)$, both the SIR coverage probability and potential throughput approach zero when $\lambda \to \infty$. For the concavity part, the proof is similar to the one provided in the Appendix D. Finally, the ASE limit follows from Appendix E and the fact that the ASE including noise is also upper-bounded by the constant given in Corollary 8. □

## V. Numerical Results and Insights

### A. Coverage Probability

We start this section by verifying our derived expressions using independent system level simulations. The simulation uniformly drops BSs in a $20 \times 20$ km$^2$ region according to the desired density. Then the SIR is evaluated for users uniformly and independently dropped in the BSs association areas in the $4 \times 4$ km$^2$ square region at the center to avoid edge effects. The results were averaged over $10^4$ runs, with 20 different users in each simulation run. Unless otherwise stated, the values of $\alpha$ are set as in to Fig. 1 and $\theta$ is set to 5 dB.

In Fig. 3, we plot the coverage probability vs. the coverage threshold $\theta$ for different BS densities, where the lower bound is based on Corollary 4. Fig. 3 shows a close match between the analytical and simulation results. Moreover, it confirms that the lower bound mimics the performance trends of the exact results and that it is tight for high densities. Fig. 3 also shows that the coverage probability depends on the BS density $\lambda$, which we already have proven analytically in the previous section. Note that increasing $\lambda$ has two significant effects on the SIR: increasing the number of nearby interfering BSs (negative effect), and decreasing the serving distance between the user and its serving BS (positive effect). These effects were proven to perfectly balance each other using the standard power-law path loss model [21] which led to the density-invariance property. However, the negative effect dominates under the proposed path loss model which leads to degradation of the coverage probability.

### B. Throughput Scaling

Although the coverage probability provides valuable insights on the network performance, it does not capture the number of BSs per unit area. Hence, we plot Fig. 4 which shows the potential throughput vs. the BS density $\lambda$. As Fig. 4 shows the potential throughput is not monotonic, and it has a maximum value after which the throughput collapses. Also, the optimal density for $\beta = 2$ agrees with the derived one in Corollary 5. This is another significant difference between the proposed model and the one used in [21] where the throughput increases at least linearly with the BS density.

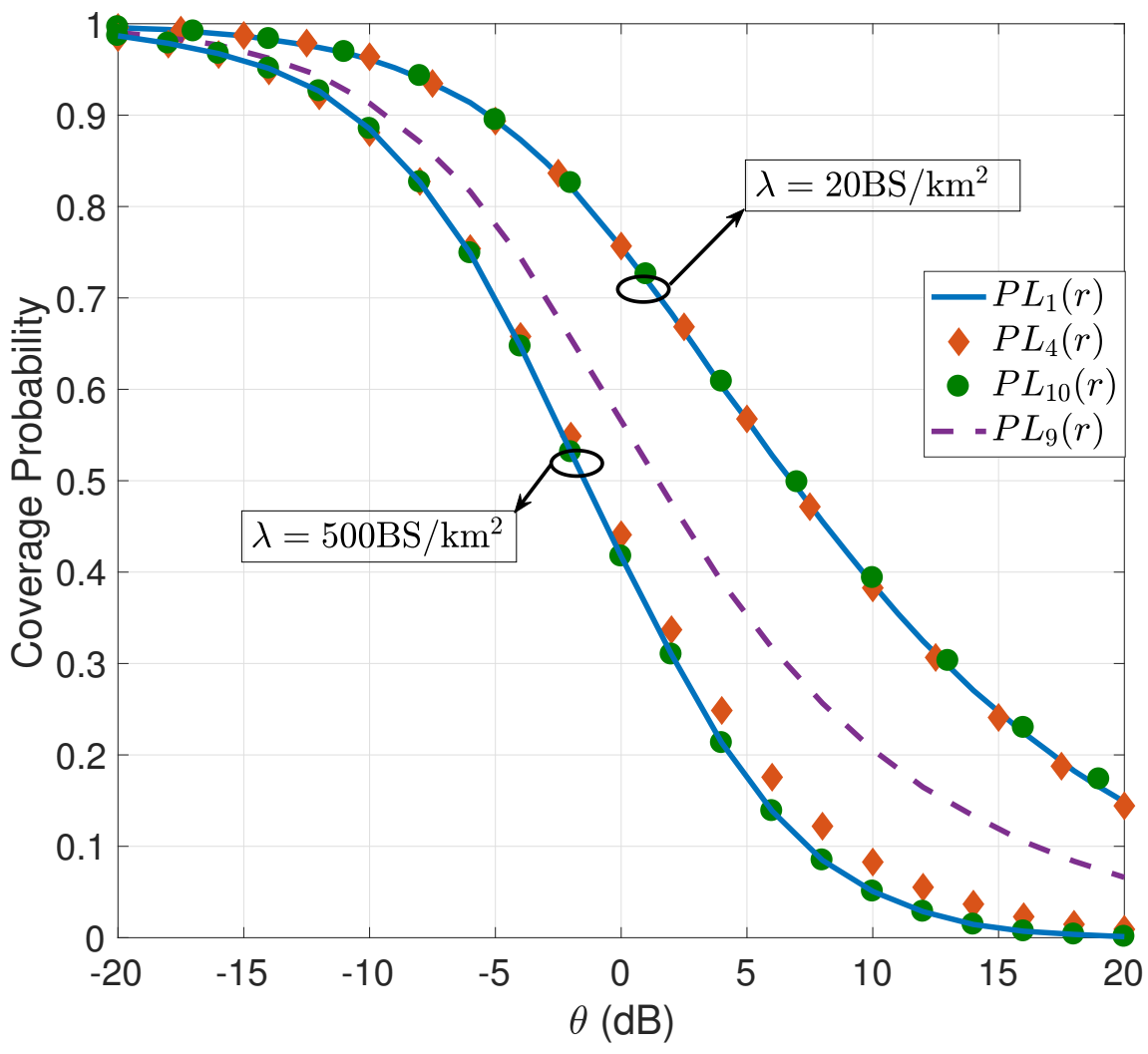

Fig. 5: The coverage probability vs. $\theta$ for different BS densities assuming different path loss models.

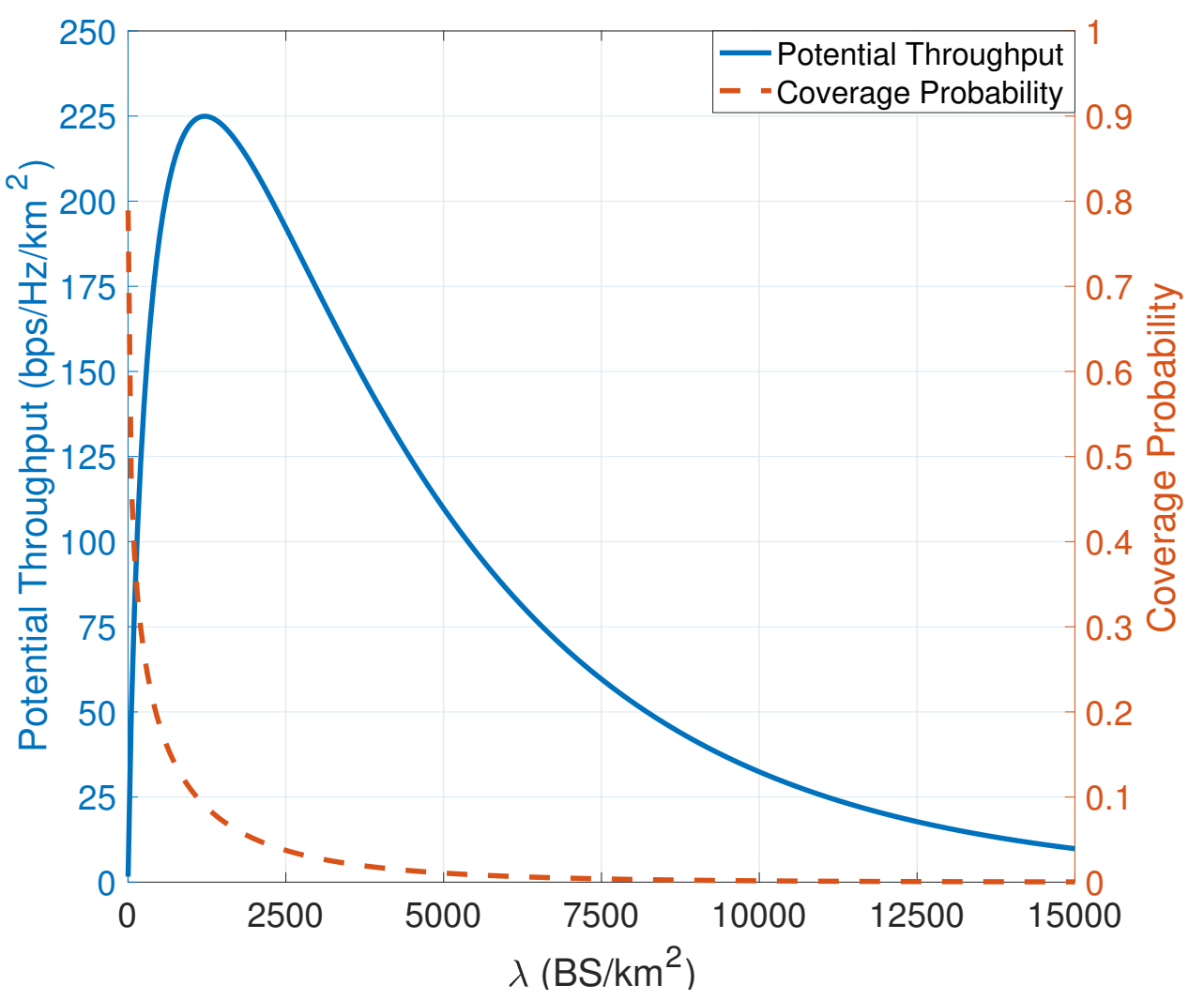

Fig. 6: The potential throughput and coverage probability vs. $\lambda$ with coverage threshold $\theta = 5$ dB.

As we discussed in Section III, the proposed path loss model is more suitable for small to medium distances. However, the performance in dense cellular networks is limited by BSs in a short distances range. To show this, we plot Fig. 5, which shows the coverage probability for different BS densities assuming different path loss models. The considered path loss models are $PL_1(r)$, $PL_4(r)$, and $PL_9(r)$ as shown in Table II with the same parameter values that lead to the best fit with the measurements. Also, we show the results for this path loss model

$$PL_{10}(r) = \begin{cases} e^{-\alpha r^{\beta}}, & r \leq 350 \ m. \\ r^{-\eta}, & r > 350 \ m. \end{cases} \tag{24}$$

In this model, the signals from BSs within the range $r \in [0, 350]$ attenuate according to the proposed model with values of $\alpha$ and $\beta$ as in Table II. Otherwise, the attenuation follows the widely-used power-law model with $\eta = 4$. As Fig. 5 shows, there is no notable difference in performance between the proposed model $PL_1(r)$ and $PL_{10}(r)$ in the dense BS scenario. Hence, it is sufficient to characterize the path loss within this range to study the performance of the network. Moreover, Fig. 5 shows the dependency of the coverage probability (the SIR distribution) on the BS density when adopting the models $PL_1(r)$, $PL_4(r)$, and $PL_{10}(r)$. However, the coverage probability is independent of the BS density when we assume the power-law model $PL_9(r)$ for small and large distances with $\eta = 4$. This observation is proved analytically in [21]. Note that there is a significant drop in the coverage probability by increasing the BS density, around 45% at $\theta = 0$ dB.

We have shown that the values $\beta = 0.5210$ and $\alpha = 0.9091$ are practical values and that we can accurately characterize the overall network performance using our proposed model. The network performance for these specific values is shown in Fig. 6. The figure shows the potential throughput and the coverage probability vs. the BS density. The coverage probability is exponentially decreasing and approaches zero for high BS density, but the potential throughput is maximized for $\lambda = 1200$ BS/km$^2$, then it collapses and reaches zero.

### C. Optimal Coverage Threshold

Note that all results in Fig. 4 and Fig. 6 hold under the assumption of fixed coverage threshold. We also consider another interesting case, where this threshold is adapted to the BS density. This case is shown in Fig. 7a, which plots the optimal coverage threshold $\theta^*$ vs. $\lambda$, where $\theta^* = \underset{\theta}{\operatorname{argmax}}\ \mathcal{R}(\lambda, \theta)$, and the optimal values were found using the grid search method. As Fig. 7a shows, densifying the network should be accompanied with lower coverage threshold, and fixing the coverage threshold for all densities will lead to performance degradation. To see this more clearly, we plot Fig. 7b which shows the optimal potential throughput $\mathcal{R}(\lambda, \theta^*)$ and the corresponding coverage probability, where the values of $\theta^*$ are the same as in Fig. 7a. The results show that the coverage probability is almost constant after $\lambda = 400$ BS/km$^2$ and the potential throughput increases with densifying the network in the plotted range. Hence, densifying the network should be accompanied with lower coverage threshold to overcome the negative effect of the increase in the network interference and achieve throughput gains. This means that we should use low rate codes, like heavily coded BPSK and/or spread spectrum techniques Otherwise, the throughput will collapse, and the network performance will degrade.

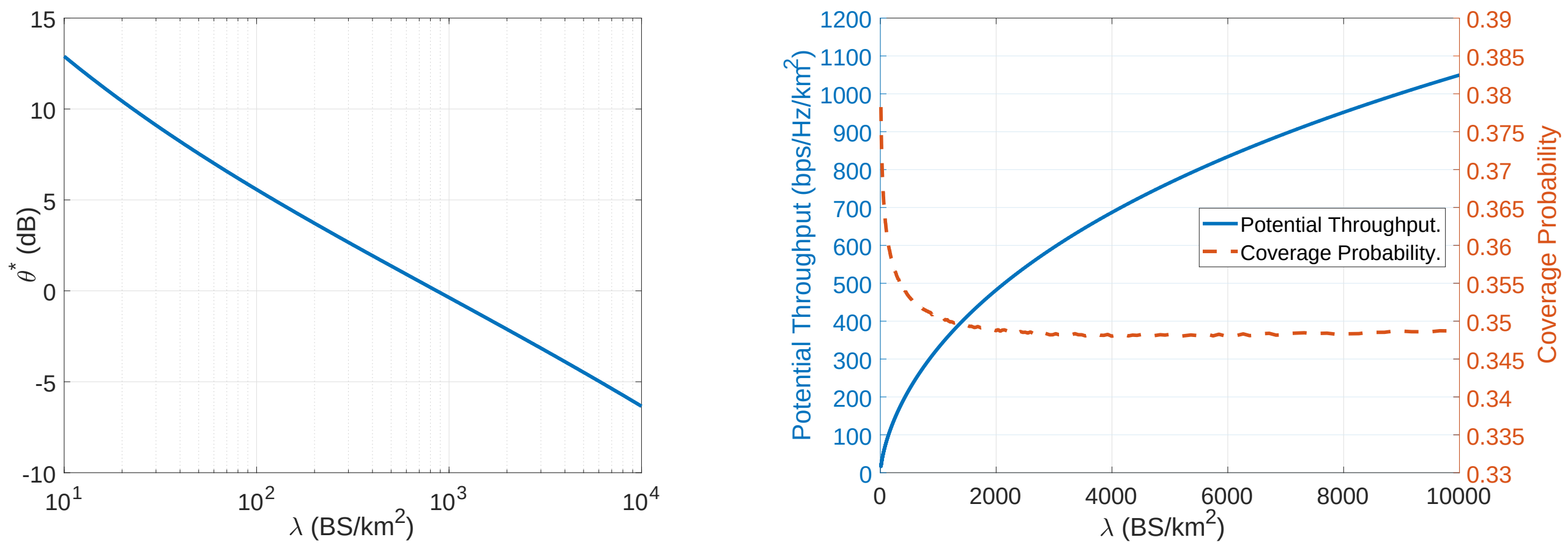

(a) The optimal SIR threshold vs. density.

(b) $\mathcal{R}(\lambda, \theta^*)$ with the corresponding $P_{cov}(\theta^*)$ vs. $\lambda$.

Fig. 7: The effect of the optimal SIR threshold for $\alpha = 1.037$ and $\beta = 0.5$.

However, it is still possible that potential throughput will eventually drop to zero after a certain BS density (larger than the shown BS range) even if the coverage threshold is tuned to $\theta^*$. Note that since the ASE is larger than $\mathcal{R}(\lambda, \theta^*)$ (see (2)), and since ASE is bounded from above by a finite constant (Corollary 8), then $\mathcal{R}(\lambda, \theta^*)$ is also bounded above by the finite constant which implies that even when we adapt $\theta$ according to the BS density, we will eventually hit a densification plateau in the best case scenario.

### D. Average Area Spectral Efficiency

We start by plotting the ASE vs. $\lambda$ as shown in Fig. 8. In Fig. 8a, we show ASE for $\beta = 2, 1, 2/3, 0.5$ along with the upper bound given in (22). Note that in all cases, the ASE converges to the upper bound as proved in Theorem 3. However, Fig. 8a does not show how fast does the ASE converge for different values. Hence, we plot Fig. 8b-d, which is a zoomed-in version of Fig. 8a for the cases of $\beta = 1, 2/3, 0.5$. As the figure shows, the ASE convergence speed decreases with increasing $\beta$.

We can also compare the potential throughput with the ASE as in Fig. (9), where we plot $\mathcal{R}(\lambda, 5dB)$ shown in Fig. 6, $\mathcal{R}(\lambda, \theta^*)$ shown in Fig. 7a, and $\mathcal{E}(\lambda)$ shown in the last figure. As expected, the results satisfy the following relation, which we have already proven in Section II.

$$\mathcal{R}(\lambda, \theta) \leq \mathcal{R}(\lambda, \theta^*) \leq \mathcal{E}(\lambda), \quad \forall \lambda > 0. \tag{25}$$

Hence, the actual network performance lies in between $\mathcal{R}(\lambda, \theta^*)$ (a pessimistic measure) and $\mathcal{E}(\lambda)$ (an optimistic measure), which means that it is always finite regardless of the chosen BS density. This achieves our main objective in this paper, which is to prove the existence of a densification plateau or possibly a collapse for this class of attenuation functions, which implies that we cannot always provide higher rates to the users by simply deploying more BSs.

## VI. Conclusion

In this work, we proposed a stretched exponential path loss model $l(r) = e^{-\alpha r^\beta}$ where $\alpha, \beta > 0$ are tunable parameters and incorporated it in stochastic geometry analysis for downlink cellular networks. We also verified the suitability of the proposed model to capture the path loss attenuation in the short to medium distance ranges by actual measurements taken the range $[5, 315]$m. We derived expressions for the coverage probability, the potential throughput, and the average area spectral efficiency and highlighted special cases where closed/compact form expressions were possible along with bounds on the considered performance metrics. Using the proposed model, we showed that the SIR distribution depends on the BS density, unlike the behavior of the standard power-law path loss model. Due to this dependence, we proved the following: 1) The coverage probability converges to zero for high BS density. 2) The potential throughput is maximized for a finite BS density then it collapses and reaches zero for high BS density. 3) The average area spectral efficiency is a non-decreasing function of the BS density and it approaches a finite constant for high BS density. Overall, we proved the existence of a densification plateau, which means that we cannot provide indefinitely higher data rate to the users by simply deploying more BSs.

Although our study focused on downlink cellular networks, the proposed model can be used to study the throughput scaling with the density in different system setups. Extensions of interest could include uplink cellular networks, ad hoc networks, MIMO systems, D2D networks, mmWave, HetNets, and interference-avoidance and suppression techniques that could fundamentally alter the SIR scaling with density.

## Acknowledgment

The authors wish to thank Prof. M. Franceschetti for sharing the measurements from [6].

## Appendix A
## Proof of Theorem 1

First we need to characterize the set of interfering BSs. Due to the closet BS association, the set of interfering BSs includes all the BSs in the network except the serving BS. With this in mind, and by following a similar approach as in [21], the LT of the interference PDF as a function of the serving distance $(r)$ can be expressed as

$$\mathcal{L}_{\mathcal{I}}(s) = \exp\left( \mathbb{E}_h\left[ -2\pi\lambda \int_r^{\infty} (1 - \exp(-she^{-\alpha z^{\beta}}))zdz \right] \right),$$

$$= \exp\left( \frac{2\pi\lambda}{\beta(-\alpha)^{\frac{2}{\beta}}} \int_0^{e^{-\alpha r^{\beta}}} \frac{\mathbb{E}_h\left[1 - \exp(-shx)\right]}{x} \ln(x)^{\frac{2-\beta}{\beta}} dx \right), \tag{26}$$

$$= \exp\left( -\frac{2\pi\lambda}{\beta\alpha^{\frac{2}{\beta}}} \int_0^{e^{-\alpha r^{\beta}}} \frac{s}{1+sx} \left(-\ln(x)\right)^{\frac{2-\beta}{\beta}} dx \right), \tag{27}$$

where (26) follows by the substitution $x = \exp(-\alpha z^{\beta})$ and (27) follows by averaging over the channel fading power, $h \sim \exp(1)$.

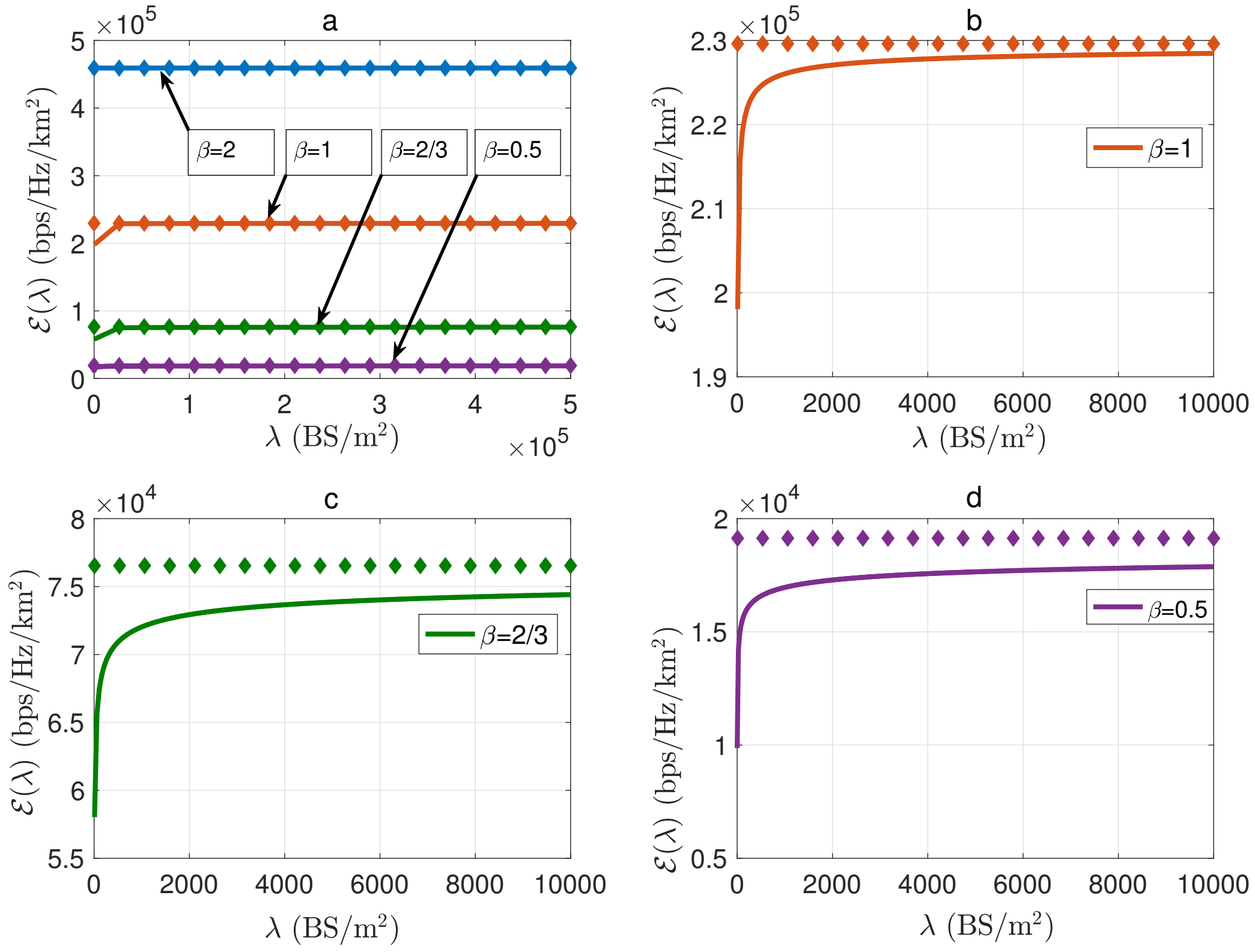

Fig. 8: The ASE vs. $\lambda$ for $\alpha = 1$ and different values of $\beta$. The upper bounds, shown as diamonds, are based on (22).

## Appendix B
## Proof of Corollary 1

For the special case of $\beta = \frac{2}{n+1}$, where $n$ is any non-negative integer, the LT in (27) can be represented by

$$\mathcal{L}_{\mathcal{I}}\left(\theta e^{\alpha r^{\beta}}\right) = \exp\left( \frac{(n+1)\pi\lambda\theta}{(-\alpha)^{n+1}} \int_0^1 \frac{\left(\ln(w) - \alpha r^{\frac{2}{n+1}}\right)^n}{1+\theta w} dw \right), \tag{28}$$

$$= \exp\left( \sum_{k=0}^{n} \binom{n}{k} (-1)^k \alpha^k r^{\frac{2k}{n+1}} \frac{(n+1)\pi\lambda\theta}{(-\alpha)^{n+1}} \int_0^1 \frac{\ln(w)^{n-k}}{1+\theta w} dw \right), \tag{29}$$

$$= \exp\left( \sum_{k=0}^{n} \binom{n}{k} \alpha^k r^{\frac{2k}{n+1}} \frac{(n+1)\pi\lambda\theta}{\alpha^{n+1}} \int_0^{\infty} \frac{-1}{e^u + \theta} (u)^{n-k} du \right), \tag{30}$$

$$= \exp\left( \sum_{k=0}^{n} \binom{n}{k} \alpha^k r^{\frac{2k}{n+1}} \frac{(n+1)\pi\lambda}{\alpha^{n+1}} (n-k)!\ \mathrm{Li}_{(n-k+1)}(-\theta) \right) \tag{31}$$

$$= \exp\left( \sum_{k=0}^{n} \lambda\ a_k(\theta)\ r^{\frac{2k}{n+1}} \right), \tag{32}$$

where (28) follows by substituting $s = \theta e^{\alpha r^{\beta}}$ and $w = x e^{\alpha r^{\beta}}$, (29) follows by the binomial expansion, (30) follows by the substitution $u = \ln(w)$, (31) by using the integral representation of the polylogarithmic function in [42], and (32) by defining $a_k(\theta) = \frac{\pi(n+1)!}{k!\alpha^{n-k+1}}\ \mathrm{Li}_{(n-k+1)}(-\theta)$. By averaging over the serving distance and defining $a_{n+1} = -\pi$ we get the desired expressions.

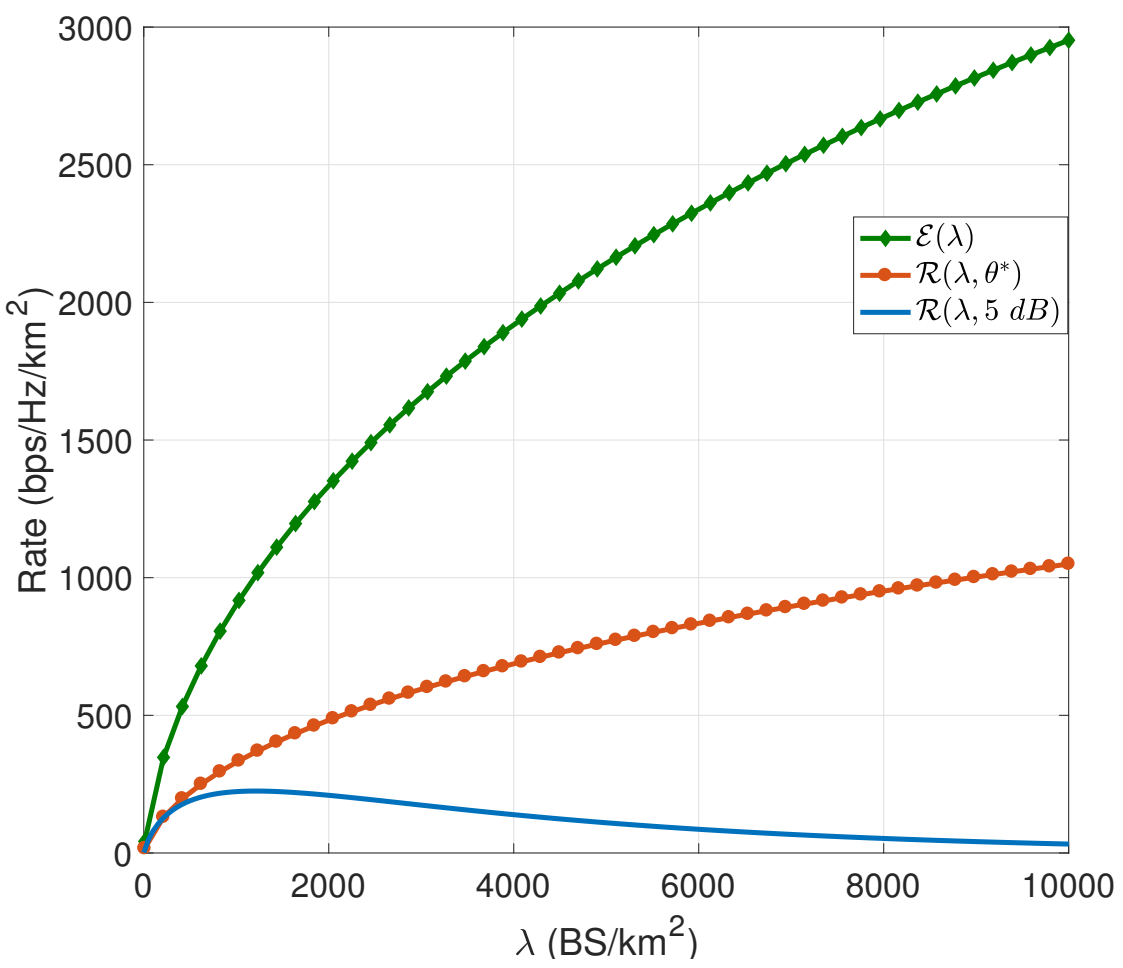

Fig. 9: The potential throughput, assuming $\theta = 5$ dB and $\theta = \theta^*$, along with the ASE vs. $\lambda$, where $\alpha = 1.037$ and $\beta = 0.5$.

## Appendix C
## Proof of Corollary 4

The key step in this proof is to show that $a_k(\theta) \leq 0, \forall k \in [0, n+1]$. For $k = n+1$, it is clear from the definition of $a_k(\theta)$ that $a_{n+1} < 0$. For $k < n+1$, $a_k(\theta)$ can be divided into two parts, $\frac{\pi(n+1)!}{k!\alpha^{n-k+1}}$ which is a positive constant, and $\mathrm{Li}_{(n-k+1)}(-\theta)$ which can be represented as this integral [42]:

$$\mathrm{Li}_{(n-k+1)}(-\theta) = \frac{-\theta}{\Gamma[n-k+1]} \int_0^\infty \frac{x^{n-k}}{e^x + \theta} dx. \quad (33)$$

Since $\theta > 0$ and $n - k + 1 \geq 0$, it is clear from (33) that $\mathrm{Li}_{(n-k+1)}(-\theta)$ is negative. Hence, $\exp(\lambda a_k(\theta) r^{\frac{2k}{n+1}}) \leq 1, \forall k \in [0, n+1]$. With this in mind, the upper-bound in (12) is found by taking the $0^{th}, n^{th}$, and $(n+1)^{th}$ terms only and (13) is found by taking the $0^{th}$ and the $(n+1)^{th}$ terms only.

To prove the upper bound in (13), we need to show that $Li_s(-\theta) \leq Li_1(-\theta) = -\ln(1+\theta), \forall s \geq 1\ \&\ \theta > 0$. The proof is as follows

$$\begin{aligned}
&Li_s(-\theta) - Li_{s-1}(-\theta) \\
&= \frac{-\theta}{\Gamma[s]} \int_0^\infty \frac{x^{s-1}}{e^x + \theta} dx - \frac{-\theta}{\Gamma[s-1]} \int_0^\infty \frac{x^{s-2}}{e^x + \theta} dx, && (34) \\
&= \frac{-\theta}{\Gamma[s]} \int_0^\infty \frac{x^{s-1}}{e^x + \theta} dx - \frac{-\theta}{\Gamma[s]} \int_0^\infty (s-1) \frac{x^{s-2}}{e^x + \theta} dx, && (35) \\
&= \frac{-\theta}{\Gamma[s]} \left( \int_0^\infty \frac{x^{s-1}}{e^x + \theta} dx - \int_0^\infty \frac{(s-1)x^{s-2}}{e^x + \theta} dx \right), && (36) \\
&= \frac{-\theta}{\Gamma[s]} \left( \int_0^\infty \frac{x^{s-1}}{e^x + \theta} dx - \int_0^\infty \frac{x^{s-1}}{e^x + \theta} \frac{e^x}{e^x + \theta} dx \right), && (37)
\end{aligned}$$

where (34) follows from the definition of the polylogarithmic function in (9), (35) by using the identity $(x\Gamma[x] = \Gamma[x+1])$, and (37) by integration by parts (it is straight forward to verify that the boundary term in the integration by parts is zero). Note that $\frac{e^x}{e^x+\theta} \leq 1, \forall x \geq 0\ \&\ \theta \geq 0$, hence the term between the parenthesis is non-negative. Since $\frac{-\theta}{\Gamma[s]} \leq 0, \forall \theta \geq 0\ \&\ s \geq 1$, then the difference in (34) is non-positive, which means that $Li_s(-\theta)$ is non-increasing w.r.t. $s$ and $Li_s(-\theta) \leq Li_1(-\theta) = -\ln(1+\theta), \forall s \geq 1\ \&\ \theta > 0$, which completes the proof of (14). The lower bound in (15) is found using Jensen's inequality.

## Appendix D
## Proof of Corollary 5

Based on Corollary 4, the coverage probability is upper bounded by $\exp\left(\lambda \frac{\pi(n+1)!}{\alpha^{n+1}} \mathrm{Li}_{(n+1)}(-\theta)\right)$, and we showed in Appendix C that $\mathrm{Li}_{(n+1)}(-\theta)$ has a negative value for all positive values of $\theta$, hence it is straightforward to show that the upper-bound approaches zero when $\lambda \to \infty$. Hence, the coverage probability approaches zero when $\lambda \to \infty$. Following a similar approach, it can be shown that the potential throughput also approaches zero when $\lambda \to \infty$.

For the log-concavity part, note that the potential throughput is given by

$$\mathcal{R}(\lambda, \theta) = \int_0^\infty 2\pi\lambda^2 r \exp\left(\sum_{k=0}^{n+1} \lambda\ a_k(\theta)\ r^{\frac{2k}{n+1}}\right) \log_2(1+\theta) dr.$$

The integrand $2\pi\lambda^2 r \exp\left(\sum_{k=0}^{n+1} \lambda\ a_k(\theta)\ r^{\frac{2k}{n+1}}\right) \log_2(1+\theta)$ is log-concave w.r.t. $\lambda$ for all values of $r$. Hence, based on [45, Section 3.4], the integration over $r$ is also a log-concave function w.r.t. $\lambda$.

## Appendix E
## Proof of Theorem 3

The proof follows by proving the following three components:

1) Prove that $\mathcal{E}(\lambda) \leq \frac{\alpha^{n+1}}{\pi(n+1)!} \log_2 e\ \forall \lambda > 0$.
2) Prove that $\lim_{\lambda \to \infty} \mathcal{E}(\lambda)$ exists and finite.
3) Prove that $\lim_{\lambda \to \infty} \mathcal{E}(\lambda) \geq \frac{\alpha^{n+1}}{\pi(n+1)!} \log_2 e$.

By combining (1) and (3) we get $\frac{\alpha^{n+1}}{\pi(n+1)!} \log_2 e \leq \lim_{\lambda \to \infty} \mathcal{E}(\lambda) \leq \frac{\alpha^{n+1}}{\pi(n+1)!} \log_2 e$. Hence $\lim_{\lambda \to \infty} \mathcal{E}(\lambda) = \frac{\alpha^{n+1}}{\pi(n+1)!} \log_2 e$. Note that (1) is showed in Corollary 8, where we proved the upper-bound and that $\mathcal{E}(\lambda)$ is log-concave w.r.t. $\lambda$. Since $\mathcal{E}(\lambda)$ is positive, upper-bounded by a finite constant, and log-concave w.r.t. $\lambda$, it follows that the limit exists and finite, which proves (2). In order to prove (3), we define $\lambda = \lambda_0 k$, where $\lambda_0 \in \mathbb{R}_+^*$ and $k \in \mathbb{N}_+^*$. Also, define $f_k = \lambda_0 k \log_2(1 + \mathrm{SINR(k)})$, and

$$\mathrm{SINR(k)} = \frac{h_0 e^{-\alpha r_0^\beta}}{\sum\limits_{r_i \in \Psi_k \setminus B(0, r_0)} e^{-\alpha r_i^\beta} h_i + N_0}, \quad (38)$$

where $h_0$ is the desired channel power gain, $\Psi_k$ is a PPP with density $k\lambda_0$, $B(x,y)$ is the disk centered at $x$ with radius $y$, $N_0$ is the average noise power and $r_0$ is the serving distance which has a PDF given by $f(r) = 2\pi k\lambda_0 r e^{-\pi k\lambda_0 r^2}$. Using these definitions, the following hold true.

$$\lim_{k\to\infty} e^{-\alpha r_0^\beta} \to 1. \;\; a.s.$$
$$\lim_{k\to\infty} \frac{N_0}{k} \to 0.$$
$$\lim_{k\to\infty} \frac{h_0}{k} \to 0. \;\; a.s. \tag{39}$$

where $a.s.$ denotes *almost sure* convergence. Note that $\Psi_k \backslash B(0,r_0)$ is the set that contains all the BSs in $\Psi_k$ except the serving BS, hence it can be represented as

$$\sum_{r_i\in\Psi_k\backslash B(0,r_0)} e^{-\alpha r_i^\beta} = \sum_{r_i\in\Psi_k} e^{-\alpha r_i^\beta} h_i - e^{-\alpha r_0^\beta} h_0, \tag{40}$$
$$= \sum_{j=1}^{k}\sum_{r_{ij}\in\Phi_j} e^{-\alpha r_{ij}^\beta} h_{ij} - e^{-\alpha r_0^\beta} h_0, \tag{41}$$

where (41) is found by exploiting the superposition property of a PPP, which means that since $\Psi_k$ has a density $k\lambda_0$, it is equivalent to the sum of $k$ i.i.d. PPPs of density $\lambda_0$, denoted by $\Phi_j,\; j\in\{1,2,...,k\}$. Hence, the limit can be represented as

$$\lim_{k\to\infty}\frac{1}{k}\sum_{r_i\in\Psi_k\backslash B(0,r_0)} e^{-\alpha r_i^\beta}$$
$$= \lim_{k\to\infty}\frac{1}{k}\sum_{j=1}^{k}\sum_{r_{ij}\in\Phi_j} e^{-\alpha r_{ij}^\beta} h_{ij} - \frac{e^{-\alpha r_0^\beta} h_0}{k}, \tag{42}$$
$$= \mathbb{E}\left[\sum_{r_{i1}\in\Phi_1} e^{-\alpha r_{ij}^\beta} h_{ij}\right] = \mathbb{E}\left[\sum_{r_{i1}\in\Phi_1} e^{-\alpha r_{ij}^\beta}\right], \tag{43}$$
$$= \int_0^\infty e^{-\alpha r^\beta} 2\pi\lambda_0 r dr = \frac{\lambda_0\pi(n+1)!}{\alpha^{n+1}}, \tag{44}$$

where (43) follows from the law of large numbers, where the empirical mean of $k$ i.i.d. RVs converges to the mean of the RV, when $k\to\infty$, and the assumption of the independence between the fading gains and the locations of the BSs and users. (44) follows directly from Campbell's theorem [18].

Next, we are interested in finding the limit of $f(k)$ as $k\to\infty$. However, we have already shown that the SINR approaches 0 for high densities, hence $\ln(1+\text{SINR}) \sim \text{SINR}$ when $k\to\infty$. So the limit of $f_k$ can be represented as

$$\lim_{k\to\infty} f_k = \log_2(e)\lim_{k\to\infty}\lambda_0 k\frac{h_0 e^{-\alpha r_0^\beta}}{\sum\limits_{r_i\in\Psi_k\backslash B(0,r_0)} e^{-\alpha r_i^\beta} h_i + N_0},$$
$$= \log_2(e)\frac{h_0\alpha^{n+1}}{\pi(n+1)!}, \tag{45}$$
$$\mathbb{E}\left[\lim_{k\to\infty} f_k\right] = \log_2(e)\frac{\alpha^{n+1}}{\pi(n+1)!}, \tag{46}$$

where (45) follows from (39) and (44). Finally, using Fatou's Lemma [46], we have $\mathbb{E}\left[\liminf_{k\to\infty} f_k\right] \le \liminf_{k\to\infty}\mathbb{E}\left[f_k\right]$, where $\liminf = \lim$ since we proved that the limits exist. Hence, according to our definition of $f_k$, we can conclude that $\mathcal{E}(\lim_{\lambda\to\infty}\lambda) \le \lim_{\lambda\to\infty}\mathcal{E}(\lambda)$ which completes the proof of Theorem 4 and the last part of Theorem 5 since we considered the SINR. Finally, since $\mathcal{E}(\lambda)$ is log-concave w.r.t. $\lambda$ and it is upper-bounded by a constant and approaches it as $\lambda\to\infty$, it follows that $\mathcal{E}(\lambda)$ is non-decreasing w.r.t. $\lambda$.

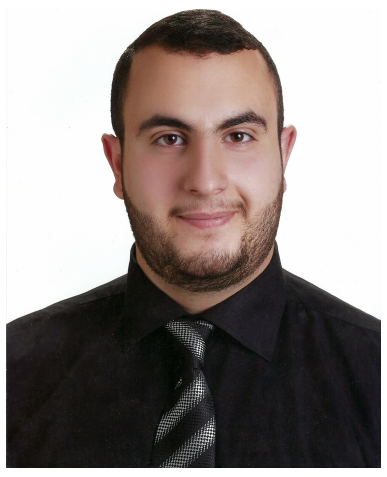

**Ahmad AlAmmouri** (S'11) received his B.Sc. degree in Electrical Engineering (with Hons.) from the University of Jordan, Amman, Jordan, in 2014 and his M.Sc. degree in Electrical Engineering from King Abdullah University of Science and Technology (KAUST), Thuwal, Saudi Arabia, in 2016. He is currently working toward the Ph.D. degree with the Department of Electrical and Computer Engineering, The University of Texas at Austin, TX, USA. His research interests include statistical modeling and performance analysis of wireless networks using stochastic geometry.

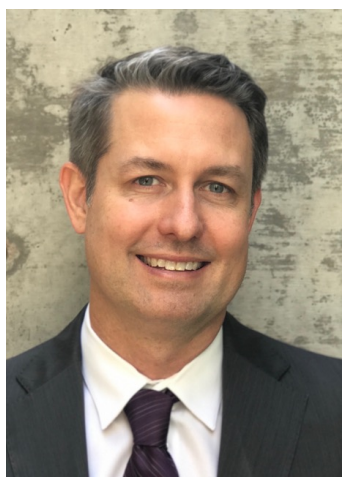

**Jeffrey Andrews** (S98, M02, SM06, F13) received the B.S. in Engineering with High Distinction from Harvey Mudd College, and the M.S. and Ph.D. in Electrical Engineering from Stanford University. He is the Cullen Trust Endowed Professor (#1) of ECE at the University of Texas at Austin. He was the Editor-in-Chief of the IEEE Transactions on Wireless Communications from 2014-2016. He developed Code Division Multiple Access systems at Qualcomm from 1995-97, and has consulted for entities including Apple, Samsung, Verizon, AT&T, the WiMAX Forum, Intel, Microsoft, Clearwire, Sprint, and NASA. He is a member of the Technical Advisory Board of Artemis Networks and GenXComm, and is co-author of the books *Fundamentals of WiMAX* (Prentice-Hall, 2007) and *Fundamentals of LTE* (Prentice-Hall, 2010).

Dr. Andrews is an ISI Highly Cited Researcher, received the National Science Foundation CAREER award in 2007 and has been co-author of fourteen best paper award recipients including the 2016 IEEE Communications Society & Information Theory Society Joint Paper Award, the 2011 and 2016 IEEE Heinrich Hertz Prize, the 2014 IEEE Stephen O. Rice Prize, and the 2014 IEEE Leonard G. Abraham Prize. He received the 2015 Terman Award, is an IEEE Fellow, and is an elected member of the Board of Governors of the IEEE Information Theory Society.

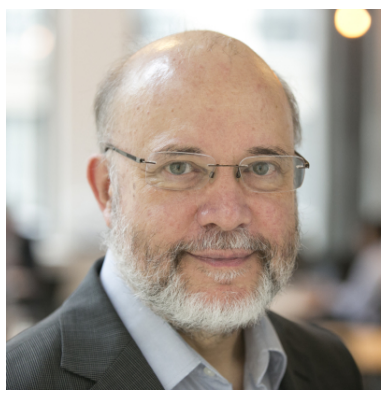

**François Baccelli** is Simons Math+X Chair in Mathematics and ECE at UT Austin. His research directions are at the interface between Applied Mathematics (probability theory, stochastic geometry, dynamical systems) and Communications (network science, information theory, wireless networks). He is coauthor of research monographs on point processes and queues (with P. Brmaud); max plus algebras and network dynamics (with G. Cohen, G. Olsder and J.P. Quadrat); stationary queuing networks (with P. Brmaud); stochastic geometry and wireless networks (with B. Blaszczyszyn). Before joining UT Austin, he held positions in France, at INRIA, Ecole Normale Suprieure and Ecole Polytechnique. He received the France Télécom Prize of the French Academy of Sciences in 2002 and the ACM Sigmetrics Achievement Award in 2014. He is a co-recipient of the 2014 Stephen O. Rice Prize and of the Leonard G. Abraham Prize Awards of the IEEE Communications Theory Society. He is a member of the French Academy of Sciences and part time researcher at INRIA.